# Air pollution in a tropical city: the relationship between wind direction and lichen bio-indicators in San José, Costa Rica

Erich Neurohr Bustamante[1], Julián Monge-Nájera[1] & María Isabel González Lutz[2]
1. Laboratorio de Ecología Urbana, UNED, 2050 San José, Costa Rica; erichneurohr@gmail.com, julianmonge@gmail.com
2. Escuela de Estadística, Universidad de Costa Rica, 2060 San José, Costa Rica; mariaisabel.gonzalezlutz@ucr.ac.cr

**Abstract:** Lichens are good bio-indicators of air pollution, but in most tropical countries there are few studies on the subject; however, in the city of San José, Costa Rica, the relationship between air pollution and lichens has been studied for decades. In this article we evaluate the hypothesis that air pollution is lower where the wind enters the urban area (Northeast) and higher where it exits San José (Southwest). We identified the urban parks with a minimum area of approximately 5 000m² and randomly selected a sample of 40 parks located along the passage of wind through the city. To measure lichen coverage, we applied a previously validated 10 x 20cm template with 50 random points to five trees per park (1.5m above ground, to the side with most lichens). Our results (years 2008 and 2009) fully agree with the generally accepted view that lichens reflect air pollution carried by circulating air masses. The practical implication is that the air enters the city relatively clean by the semi-rural and economically middle class area of Coronado, and leaves through the developed neighborhoods of Escazú and Santa Ana with a significant amount of pollutants. In the dry season, the live lichen coverage of this tropical city was lower than in the May to December rainy season, a pattern that contrasts with temperate habitats; but regardless of the season, pollution follows the pattern of wind movement through the city. Rev. Biol. Trop. 59 (2): 899-905. Epub 2011 June 01.

Urban ecology research is a relatively new field, and urban ecologists have mostly done observational studies rather than formal research linking cause and effect (Shochat *et al*. 2006). A key aspect of urban ecology is the flow of atmospheric gases, particularly contaminants that affect human health (Stapper 2004, Gombert *et al*. 2006, Leak *et al*. 2008, Policnik *et al*. 2008). Lichens are good bioindicators of "air quality" and their communities integrally reflect pollution levels (Cristofolini *et al*. 2008, Rojas-Fernández *et al*. 2008, Lijteroff *et al*. 2009).

A recent book length review of lichens and monitoring (Nimis *et al*. 2002) shows that lichens are now widely used bio-indicators in several countries of Europe and North America. This trend may be particularly important for tropical countries where funds for ecological monitoring hardware are scarce or non-existent. Detailed protocols have been developed for a variety of purposes that go from monitoring specific chemical compounds to helping forest conservation on the basis of patterns shown by lichen communities (Nimis *et al*. 2002, Monnet *et al*. 2005, Nali *et al*. 2007).

The mechanisms by which contaminants affect lichens include electrical conductivity, which is reduced when the cell membranes of the thallus are degraded by air pollution (Garty *et al*. 2000). Antioxidant parameters of lichens can serve as indicators of rapid response to pollution stress (Weissman 2006). Standard methods have also been refined in the last decade for better determining the accumulation of various heavy metals on lichens, as in *Evernia prunastri* (Nimis *et al*. 2002, Conti *et al*. 2004).

Another recent global trend is the identification, modeling and recording on digital maps of changes in the lichen vegetation from urban and rural areas (Policnik *et al*. 2008). For example, digital models of lichen recolonization for the Ruhr Valley in Germany, show that there was little pollution in the 15th century, heavy pollution in the 20th century and again less pollution in the 21st century, thanks to recent interest in the recovery and conservation of healthy living conditions in cities and towns (Kricke & Beige 2004). Recently, digital maps have also proved appropriate to record the effects of air pollution on the biodiversity of lichens in Italy (Loppi 2004).

Atmospheric pollution in the vicinity of a tin and lead industry clearly correlated with wind patterns when studied on the lichen *Canoparmelia texana* (Leonardo *et al*. 2010).

In London and surrounding counties, pollution carried by wind explains the occurrence of two central regions with few lichens and no bryophytes, a surrounding region with a more diverse flora, including a high cover of nitrophyte-related lichens, and an outer region characterized by species absent from central London, including acidophyte lichens (Larsena *et al*. 2007).

In the tropics, Latin American scientists have a long trajectory of research on the relationship between lichens and air pollution (Méndez & Fournier 1980, Monge-Nájera *et al*. 2002a, Bedregal *et al*. 2005), but this work has not received the attention that it deserves; for example, there are no Latin American authors in the book edited by Nimis *et al*. (2002), not even for the chapter on tropical lichens.

Costa Rica has had an air-monitoring program based on lichens for several decades. For example, an early study assessed air pollution in its capital city, San José, using three methods: the overall coverage of urban lichens on tree trunks, the changes in lichens transplanted from low-pollution areas near the city, and counts of living cells versus dead cells in lichens exposed for long periods to urban air pollution (Méndez 1977, Méndez & Fournier 1980). A decade later, using transplanted temperate lichens in a tropical habitat, Grüninger & Monge-Nájera (1988) tested the European standard bio-monitoring lichen, *Hypogymnia physodes*, in San José.

In the 21st Century, Costa Rican lichenology includes a long term study of air pollution (Monge-Nájera *et al*. 2002a) that is now in its 35[th] year and the development of a new method for determining pollution using lichens with half the effort required by previous methods (Monge-Nájera *et al*. 2002b).

In this article we evaluate the hypothesis that air pollution is lower in areas where the air mass enters the Costa Rican capital (northeast) and higher where it exits (southwest).

## MATERIALS AND METHODS

**Sampling sites:** We studied the lichen cover on tree trunks in San José, Costa Rica, in a rectangular plot covering the area where the air enters the city (Guadalupe, Northeast) to the area where it exits (Escazú, Southwest; Lambert North Costa Rica: 527 850 East and 212 670 North, Fig. 1). The width of the plot was enough to obtain the desired population of about 200 parks.

We used public maps to find all urban parks that were large enough to have several trees with lichens (the maximal size was near 40 000m$^2$). To minimize the ecological edge effect, we limited a minimum area of approximately 5 000m$^2$. From the population of 200 parks, we selected a sample of 40 parks with a random sample generator (www.random.org).

**Measurements:** Following Monge-Nájera *et al*. (2002b), in each park we identified the five trees with highest lichen coverage on their trunks. We discarded those that did not have enough space for the template and the ones with contorted trunks (trunks must be straight in the first 2m above ground because the flow of rainwater affects lichen cover) (Monge-Nájera *et al*. 2002a,b). The five parks that had no suitable trees were replaced by others, also randomly, to reach the desired sample of 40 parks.

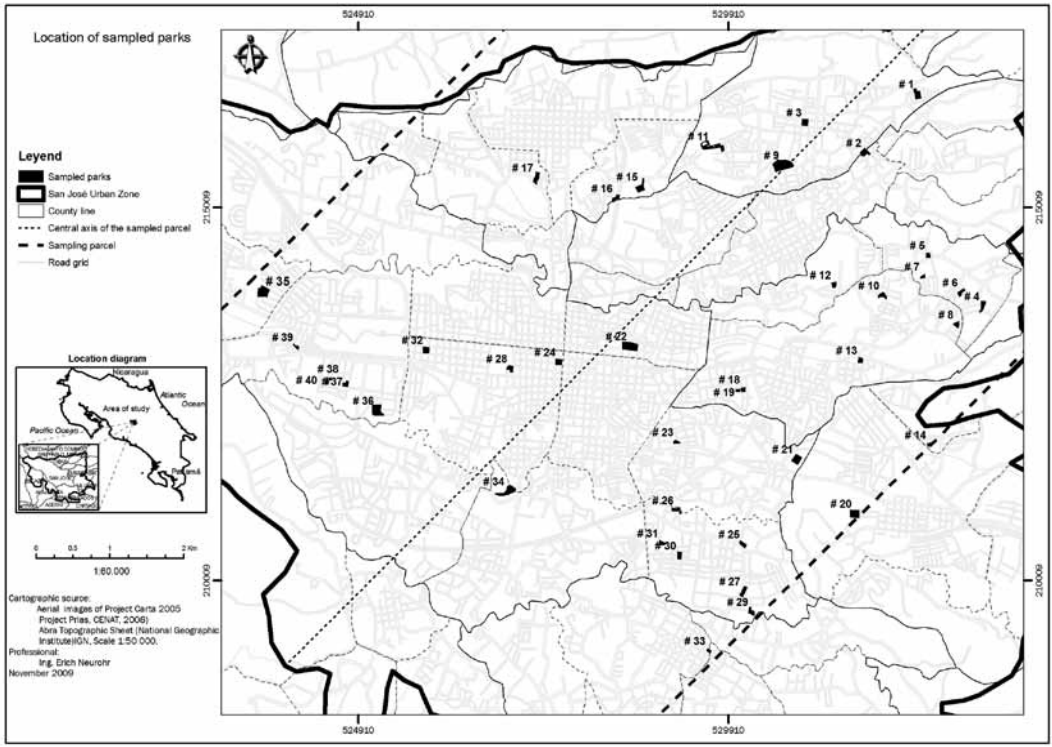

**Fig. 1.** Sampled parks (numbered) in San José, Costa Rica, 2009.

We used a 10 x 20cm template with 50 random points, applying it 1.5m above ground to the side with greater lichen coverage; a detailed description of the method and its validation are given elsewhere (Monge-Nájera *et al*. 2002b).

**Lichen species:** In temperate countries, the species of lichens used in the surveys are usually considered individually because air pollution affects not only lichen coverage, but also species richness and composition (Nimis *et al*. 2002). Since identifying tropical lichen species is time consuming, and in the case of San José, unreliable for the lack of adequate taxonomic studies (Umaña & Sipman 2002), we decided to include only the overall coverage of foliose lichens, which is an easy and reliable method for bio-monitoring air pollution (Monge-Nájera *et al*. 2002b).

**Sampling dates:** In contrast with temperate lichens, some tropical species grow more during moist periods, and they are expected to be more affected by pollution during the dry season when they are stressed but active (Monge-Nájera *et al*. 2002a, Umaña & Sipman 2002). For this reason, we sampled the end of the 2008 rainy season and the end of the 2009 dry season. To reduce sampling error, we repeated the individual trees, which for this purpose were marked with numbered metal pins.

In order to analyze the trend of lichen reaction to pollution (as reflected by coverage on the trunks) in relation with park location along the wind path, we used the following adjusted multiple regression model:

$$Y = \beta_0 + \beta_1 X_1 + \beta_2 X_2 + \varepsilon$$

Whereas:
$X_1$ = 1 for the rainy season and -1 for the dry season
$X_2$ = distance (km) from a reference line at air's entry point
Y = lichen cover on trunks (percentage)

This regression has the advantage of considering any "statistical noise" that season might have on the variable that we wanted to measure: the effect of distance from the source of pollution, in this case, smog from the city vehicles, which are the main source of air pollution in San José (Brighigna *et al*. 2002, Monge-Nájera *et al*. 2002a).

RESULTS

The phorophites belonged to these species: *Acnistus arborescens, Bauhinia purpurea, Callistemon viminales, Cassia fistula, Cedrela odorata, Chrysophyllum cainito, Citrus sinensis, Cojoba arborea, Cupressus lusitanica, Delonix regia, Enterolobium cyclocarpum, Eriobotrya japonica, Erythrina berteroana, Eucaliptus sp, Ficus benjamina, Fraxinius udhei, Hymenaea courbaril, Jacaranda mimosaefolia, Macadamia intergrifolia, Mangifera indica, Mauria heterophylla, Persea caerulea, Phoenix canariensis, Pinus* sp., *Quercus costaricensis, Spathodea campanulata, Tabebuia rosea, Terminalia catappa, Trichilia havanensis* and *Veitchia arecina*.

The distance and the season explained 13.8% of the changes in lichen coverage on the tree trunks of the city ($r^2$=0.138). This percentage can be considered a very good correlation, especially if one considers the high local variability in lichen coverage within the sampled area (Fig. 2).

Independent of the season, coverage was reduced from northeast to southwest, on average, 1.34 percentage points per kilometer. Therefore slightly polluted parks predominated in the Northeast (wind input to the city) and heavily contaminated parks in the Southwest (Fig. 2), where the air mass leaves the city after collecting a load of pollutants from vehicle exhausts.

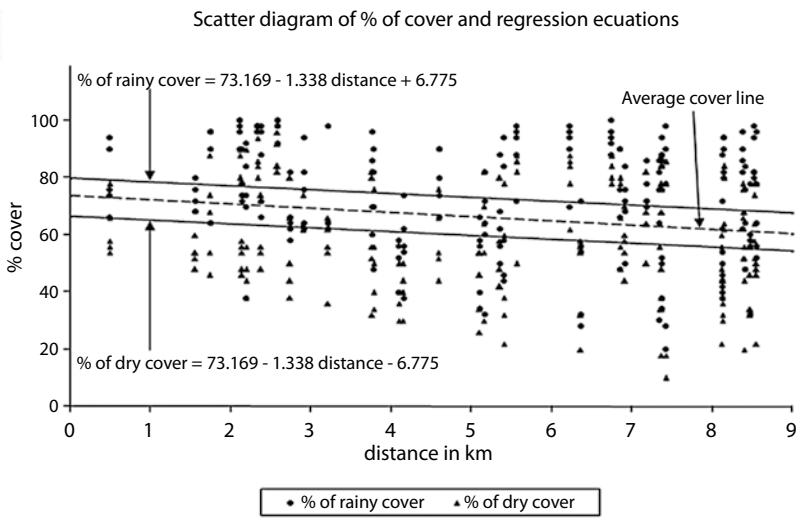

**Fig. 2.** Dispersion of lichen cover with regression equations, whereas rainy = 1 for rainy season and -1 for dry season. P values: intercept <0.0001, distance 0.0007, season <0.0001.

Regardless of the distance from the wind's point of entry, during the dry season the lichen coverage was 6.78 percentage points lower than the mean (Fig. 2).

## DISCUSSION

Few recent authors have denied that the wind carries atmospheric pollutants and that this is reflected in the lichens´ cover. One of those exceptions is the team led by Marques *et al*. (2009), which reported that when *Parmelia sulcata* transplants were placed to allow different routes of air influx, there were no clear wind-directional effects on element accumulation. Nevertheless, they warned that their results could not be generalized to other lichen species.

On the contrary, the association between lichens and wind has been found in a variety of environments for several decades (Nimis *et al*. 2002). For example, in the grazing land and fields of Lincolnshire, England, where there has been a limited urban and industrial development, when compared with sources of pollution some distance from it (Seaward 1973).

The pollutants may not travel long distances when carried by the wind: Vanadium released to the atmosphere by a power plant using heavy oil as fuel was carried by the wind towards adjacent areas, accumulating in the lichen *H. physodes* mostly at less than 1km from the power plant (lichens collected 50km from the plant contained less than 2ppm of Vanadium; Nygårda & Harjua 1983).

Our results in Costa Rica fully agree with the generally accepted view that relative lichen cover on tree trunks reflects air pollution carried by circulating air masses (Larsena *et al*. 2007). In San José, the air collects vehicle exhaust pollutants as it moves through the city, leaving it at the Southwest area with a heavier load of contaminants, in accordance with the hypothesis proposed by Monge-Nájera *et al*. (2002a) and tested in this study.

In Europe and other temperate regions, lichens are more affected by pollution in the rainy season, possibly because they are metabolically active and because it is the time when pollutants from domestic heating and others increase (Loppi 2010 pers. comm.). In San José, where neither heating nor air conditioning is necessary and there is little industrial development, the main source of air pollution is vehicle exhausts (Monge-Nájera *et al*. 2002a).

Tropical lichens can grow rapidly (Umaña & Sipman 2002) and this could explain the apparent 13.56% growth that we found a few months after the first sampling, but the growth may not be real, because this value is similar than the 11% error level of the sampling method (Monge-Nájera *et al*. 2002b).

Regardless of the season, lichens reflected the passage of an air mass that enters San José with lower pollution levels and leaves the city with a high burden of contaminants. The practical implication is that the air enters the city relatively clean by the semi-rural and economically middle class area of Coronado, and reaches the developed neighborhoods of Escazú and Santa Ana with a significant amount of pollutants.

Future studies of tropical urban lichens could explore if lichen cover actually is lower in the dry season, a result that differs from reports from temperate regions and the effect that climate change may have on tropical lichens, as done for other regions by Insarov *et al*. (1999).

## ACKNOWLEDGMENTS

We thank Ifigenia Bustamante G. for her assistance in field work and for the identification of trees, Juan Navarro R. for his help in locating the parks, and Stefano Loppi (Dipartimento di Scienze Ambientali, Universita di Siena, Italy), Harold Arias Leclaire (UNED, Costa Rica), Liis Marmor (Institute of Ecology and Earth Sciences, University of Tartu, Estonia), Víctor H. Méndez (UNED, Costa Rica) and Juri Nascimbene (Dipartimento di Biologia, Università di Trieste, Italy) for valuable suggestions to improve an earlier draft.

## RESUMEN

Los líquenes constituyen un buen bioindicador para estudiar la "salud de la atmósfera", pero en los países tropicales hay pocos estudios sobre el tema, aunque para la ciudad de San José existen algunos estudios sobre la relación entre tráfico vehicular y contaminación atmosférica. En este artículo evaluamos la hipótesis de que la contaminación atmosférica es menor en las áreas por donde ingresan los vientos a la zona urbana de San José (noreste) y mayor a su salida (suroeste), para las épocas seca y lluviosa. Para obtener parques urbanos con tamaño suficiente para albergar varios árboles con líquenes, se identificaron los parques con un área aproximada mínima de 5 000m². Seleccionamos aleatoriamente una muestra de 40 parques ubicados a lo largo del paso del viento por la ciudad. Para medir el porcentaje de cobertura, en cada parque aplicamos a cinco árboles una plantilla de 10 x 20cm con 50 puntos al azar, a 1.5m sobre el suelo y del lado con mayor cobertura de líquenes. Nuestros resultados coinciden con la opinión generalmente aceptada de que los líquenes reflejan la contaminación transportada por la circulación de masas de aire. La consecuencia práctica es que el aire entra en la ciudad relativamente limpia por el área semi-rural de clase media de Coronado, y llega a los barrios desarrollados de Escazú y Santa Ana con una cantidad importante de contaminantes. En la época seca la cobertura de líquenes vivos fue menor que en la estación lluviosa (que va de mayo a diciembre), pero independientemente de la época, la contaminación sigue el patrón de viento de la ciudad.

**Palabras clave:** monitoreo ambiental, dispersión de la contaminación, ciudad tropical, bioindicadores, plantas no vasculares.